\documentclass{jfm}
\usepackage{graphicx}
\usepackage{epstopdf, epsfig}
\usepackage{blindtext}
\usepackage{amsmath}
\usepackage{mathtools}
\usepackage[colorlinks,citecolor = blue, linkcolor=red,hyperindex,CJKbookmarks]{hyperref}

\shorttitle{Caf\'e Latte: Spontaneous layer formation}
\shortauthor{K. L. Chong \textit{et al.}}
\title{Caf\'e Latte: Spontaneous layer formation in laterally cooled double diffusive convection}
\author{Kai Leong Chong\aff{1}\corresp{\email{k.l.chong@utwente.nl}},
      Rui Yang\aff{1},
      Qi Wang\aff{1},
      Roberto Verzicco\aff{1,2,3}
      \and Detlef Lohse\aff{1,4}\corresp{\email{d.lohse@utwente.nl}}
 }
\affiliation{
\aff{1}Physics of Fluids Group, Max Planck Center for Complex Fluid Dynamics, MESA+ Institute and J.M.Burgers Center for Fluid Dynamics, University of Twente, P.O. Box 217, 7500 AE Enschede, The Netherlands
\aff{2}Dipartimento di Ingegneria Industriale, University of Rome `Tor Vergata', Via del Politecnico 1, Roma 00133, Italy
\aff{3}Gran Sasso Science Institute - Viale F. Crispi, 7 67100 L'Aquila, Italy
\aff{4}Max Planck Institute for Dynamics and Self-Organisation, 37077 G\"ottingen, Germany}
\begin{document}

\maketitle

\begin{abstract}
In the preparation of Caf\'e Latte, spectacular layer formation can occur between the expresso shot in a glass of milk and the milk itself. Xue \textit{et al.} (Nat. Commun., vol. 8, 2017, pp. 1--6) showed that the injection velocity of expresso determines the depth of coffee-milk mixture. After a while when a stable stratification forms in the mixture, the layering process can be modelled as a double diffusive convection system with a stably-stratified coffee-milk mixture cooled from the side. More specifically, we perform (two-dimensional) direct numerical simulations of laterally cooled double diffusive convection for a wide parameter range, where the convective flow is driven by a lateral temperature gradient while stabilized by a vertical concentration gradient. Depending on the strength of stabilization as compared to the thermal driving, the system exhibits different flow regimes. When the thermal driving force dominates over the stabilizing force, the flow behaves like vertical convection in which a large-scale circulation develops. However, with increasing strength of the stabilizing force, a meta-stable layered regime emerges. Initially, several vertically-stacked convection rolls develop, and these well-mixed layers are separated by sharp interfaces with large concentration gradients. The initial thickness of these emerging layers can be estimated by balancing the work exerted by thermal driving and the required potential energy to bring fluid out of its equilibrium position in the stably stratified fluid. In the layered regime, we further observe successive layer merging, and eventually only a single convection roll remains. We elucidate the following merging mechanism: As weakened circulation leads to accumulation of hot fluid adjacent to the hot sidewall, larger buoyancy forces associated with hotter fluid eventually break the layer interface. Then two layers merge into a larger layer, and circulation establishes again within the merged structure.
\end{abstract}


\section{Introduction}
Layered patterns are striking features in double diffusive convection (DDC), where the fluid density depends on two scalars with different diffusivities \citep{Turner1974,huppert1981double,Schmitt1994,Radko2013,Garaud2018a}. A typical example of layer formation is found in the Ocean, where seawater density is affected by temperature and salinity. As a result of double diffusion, thermohaline staircases are found in different regions of the Ocean, such as a salt-finger regime in (sub-)tropic regions \citep{Schmitt2005,Johnson2009,yang2019multiple} and a diffusive regime in high-latitude regions \citep{Kelley2003,Timmermans2008,Sommer2013}.

An intriguing daily example of layered pattern can be found in Caf\'e Latte, where the corresponding laboratory experiments have recently been conducted by \cite{xue2017laboratory}. When a shot of espresso (lower-density) is poured into a glass of milk (higher-density), the system cools down from the side since it loses heat to the ambient through the sidewall, pronounced layers form in the mixture, rather than a mixed-up solution as one may expect. \cite{xue2017laboratory} showed that the injection velocity determines the depth of milk being mixed with expresso. After a while when a stably-stratified zone forms in the mixture, the layering process is governed by double diffusion: The temperature difference between the hotter bulk and the colder sidewall fluid layer implies a horizontal thermal driving, whereas a stabilizing vertical concentration gradient exists in the coffee-milk mixture. 

Examples in other physical systems also illustrate the importance of horizontal thermal driving to layer formation in a stably-stratified fluid. For instance, when sedimenting suspensions of colloidal particles are subjected to a horizontal temperature gradient, the initially uniform suspension will also develop multiple layers \citep{mendenhall1923stratified}. Also, one can observe layering when ice blocks melt in salty liquid, building up a salinity gradient \citep{huppert1980ice}.

To numerically study layer formation in DDC systems with both vertical and lateral gradients, here we pick, inspired by Caf\'e Latte, laterally cooled double diffusive convection with a concentration gradient in vertical direction. In this set-up, the temperature gradient is imposed horizontally, whereas the vertical concentration gradient is stabilizing. In pioneering experimental and theoretical work of laterally cooled DDC, \cite{thorpe1969effect} showed the successive growth of layers in a stratified brine solution heated from one side. They further conducted linear stability analysis to find the onset criteria of layers. Their pioneering paper motivated further experimental and numerical work focusing on how the layers form \citep{chen1971stability,wirtz1972physical,lee1991double}. Also, salinity and heat fluxes were studied extensively in laterally cooled DDC, because it is relevant to the high-latitude Ocean being affected by melting icebergs \citep{jacobs1981thermohaline, huppert1980ice, gayen2016simulation}. Moreover, layer merging in DDC is also an important issue because it influences the fluxes across the layer interface \citep{tanny1988dynamics, chen1997salt}. 

Previous simulations (mostly in $20$th century) on DDC had severe CPU-time limitation on the parameter range and on collecting long enough time series of layer evolution. Thanks to the full temperature and velocity information obtained from present numerical simulations, the extension of the parameter space and the possibility to run very long simulations, the layer formation and properties can now be understood in much more detail.

In this work, we study laterally cooled DDC over a wide range in parameter space, namely three decades of temperature Rayleigh number $Ra_T$ and four decades of density ratio $\Lambda$. We begin with the description of the governing equations and the setup in Section \ref{sec:num}. Then we examine the flow morphologies and show the layer formation in Section \ref{sec:morph}. In Section \ref{sec:height}, we can estimate the thickness of the initially formed layers from an energy balance. We further elucidate the mechanism of layer merging in Section \ref{sec:mech}. Finally, conclusions are given in Section \ref{sec:conc}.
\section{Numerical method and set-ups}\label{sec:num}
We consider a two-dimensional rectangular box of width $W$ and height $H$. The left/right wall has high/low temperature, and there is no salinity flux through the lateral boundaries. The top/bottom wall has low/high salinity and is adiabatic to temperature. No-slip velocity boundary conditions are used on all the walls.
We apply the Oberbeck-Boussinesq (OB) approximation, such that the fluid density depends linearly on temperature $\tilde{T}$ and a scalar $\tilde{S}$: $\tilde{\rho} ( \tilde{T} , \tilde{S} ) = \tilde{\rho} _ { 0 } \left[ 1 - \beta _ {T} \left( \tilde{T} - \tilde{T} _ { 0 } \right) + \beta _ {S} \left( \tilde{S} - \tilde{S} _ { 0 } \right) \right]$. Here, $\tilde{\rho}_0$, $\tilde{T}_0$, $\tilde{S}_0$ represent the reference density, temperature and concentration, respectively. $\beta_T$ and $\beta_S$ are the thermal and solutal expansion coefficients. The governing equations are nondimensionalized by normalizing lengths by $H$, velocities by the free-fall velocity $U = \sqrt { g \beta _ {T } | \Delta_T | H }$, temperatures by $\Delta_T$ (the temperature difference between the sidewalls) and concentrations by $\Delta_S$ (the concentration difference between the top and bottom plates):
\begin{equation} \label{eq:mom}
  { \partial _ { t } u _ { i } + u _ { j } \partial _ { j } u _ { i } = - \partial _ { i } p + \sqrt { \frac { P r _ { T } } { R a _ { T } } } \partial _ { j } \partial _ { j } u _ { i } + \left( T - \Lambda S \right) \delta _ { i z } },
\end{equation}
\begin{equation}
 { \partial _ { t } T + u _ { i } \partial _ { i } T = \frac { 1 } { \sqrt { R a _ { T } P r _ { T } }} \partial _ { j } ^ { 2 } T },
\end{equation}
\begin{equation}\label{eq:concentration}
 { \partial _ { t } S + u _ { i } \partial _ { i } S =\frac { 1 } { Le \sqrt { R a _ { T } P r _ { T } } } \partial _ { j } ^ { 2 } S },
\end{equation}
\begin{equation}\label{eq:incom}
\partial _ { i } u _ { i } = 0.
\end{equation}
Here, $u_i$ are the velocity components, $p$ the kinematic pressure, $T$ the temperature and $S$ the concentration, all now non-dimensional. $\delta_{iz}$ denotes the Kronecker delta and $g$ the gravitational acceleration. The five dimensionless control parameters are the aspect ratio $\Gamma$, the thermal Rayleigh $Ra_T$ and the Prandtl $Pr_T$ number for the temperature, the Lewis number $Le$, and the density ratio $\Lambda$, defined as:
\begin{align}
 \Gamma &= W/H, \quad R a _ { T } = \frac { g \beta _ { T } H ^ { 3 } \Delta _ { T } } { \kappa _ { T } \nu } , {\quad  Pr } _ { T } = \frac { \nu } { \kappa _ { T } }, \\
 L e &= \kappa _ { T } / \kappa _ { S } =  { Pr } _ { S } P r _ { T } ^ { - 1 }, \quad  \Lambda = \left( \beta _ { S } \Delta _ { S } \right) / \left( \beta _ { T } \Delta _ { T } \right) = R a _ { S } R a _ { T } ^ { - 1 }L e  ^ { - 1 },
\end{align}
\noindent where $R a _ { S } = { g \beta _ { S } H ^ { 3 } \Delta _ { S } } /{ (\kappa _ { S } \nu) }$ is the concentration Rayleigh number and $ { Pr } _ { S } = { \nu } /{ \kappa _ { S } }$ the concentration Prandtl number. $\nu$, $\kappa_T$ and $\kappa_S$ are the kinematic viscosity, the thermal diffusivity, and the solutal diffusivity, respectively. 
 $\Lambda$ measures the relative strength of the buoyancy force induced by the stabilizing concentration difference to that induced by the destabilizing temperature difference. The three key response parameters of the system are the two scalar fluxes and the flow velocity, which are measured by the two Nusselt numbers and the Reynolds number:
\begin{align}
 Nu_T &= \sqrt{Ra_T Pr_T}\langle u_x T\rangle _{z,t} - \langle\partial_x T\rangle_{z,t},\\
 Nu_S &= \sqrt{Ra_S Pr_S}\langle u_z S\rangle_{x,t} - \langle\partial_z S\rangle_{x,t},\\
 Re   &= \sqrt{Ra_T/Pr_T}\sqrt{\langle\boldsymbol{u}^2\rangle}.
\end{align}
Here $\langle . \rangle _{x,t}$/$\langle . \rangle_{z,t}$ represents the average over time and the horizontal/vertical plane. In this work, we calculate $Nu_T$ by  temperature gradients at the two sidewalls and $Nu_S$ by concentration gradients at the top and bottom plates. $\sqrt{\langle\boldsymbol{u}^2\rangle}$ is the root-mean-square value of the velocity magnitude, calculated over the entire domain.

Equations (\ref{eq:mom})--(\ref{eq:incom}) are solved by a second-order finite difference scheme using a fractional-step procedure and advanced in time by a low-storage third-order Runge-Kutta scheme \citep{verzicco1996finite,van2015pencil}. We use fixed aspect ratio $\Gamma=0.5$.  Our simulations cover the range $10^6\leq Ra_T\leq 10^9$, $10^{-2}\leq \Lambda\leq 10^2$ with $Pr_T = 1$ and $Pr_S = 100$ (corresponding to a Lewis number $Le$=100 which is large enough to demonstrate the layer formation). The large $Le$ implies that the resolution for the concentration is more demanding than that for the temperature, and thus a multiple resolutions strategy is employed \citep{Ostilla-Monico2015}, and such strategy had been already used for DDC simulations \citep{Yang2015}. Specifically, for the case of $Ra_T=10^9$, we use $432^2$ for the base mesh and $1296^2$ for the refined mesh. 

We are aware of the limitations of 2D DDC simulations as compared to 3D ones. However, at least for large $Pr\geq 1$, qualitatively the results for 2D \& 3D are very similar \citep{van2013comparison,chong2020subcritical} and we aim more at elucidating the physical processes originating the layering rather than detailing a specific case. Only by restricting us to 2D, we can explore a large region of the parameter space. 
\begin{figure}
\centering
\centerline{\includegraphics[width=0.65\textwidth]{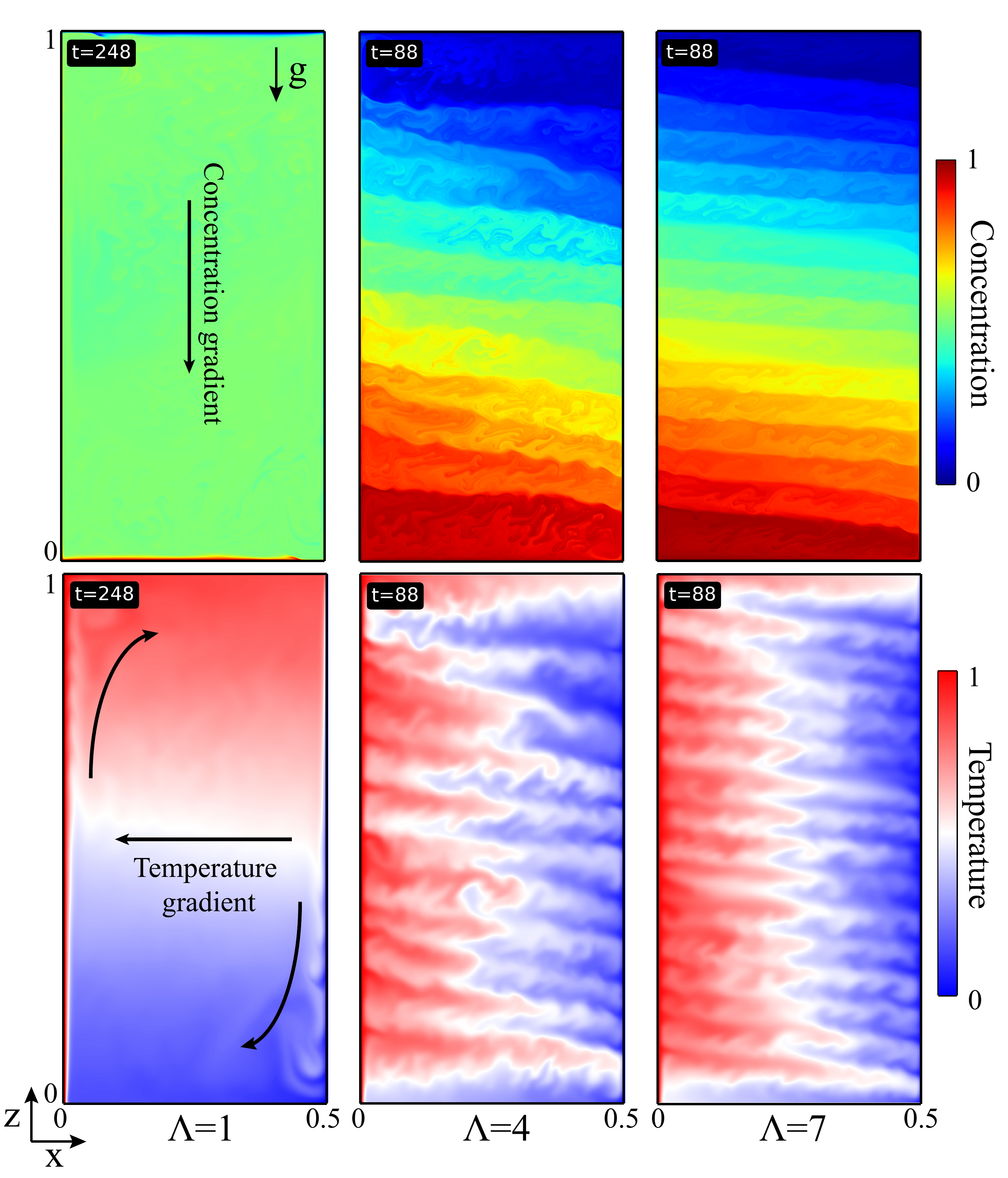}}
\caption{Snapshots of the concentration (upper row) and temperature fields (lower row) for different $\Lambda=1,4$ and $7$ (from left to right) with $Ra_T=10^9$ and $Le=100$. The global temperature difference is imposed laterally while the global concentration difference is along the vertical direction to stabilize the flow. Layered structures emerge at large enough density ratio $\Lambda$, which contrasts the domain-filling circulation (sketched by arrows) observed at $\Lambda=1$. Corresponding movies are shown in the Supplementary Materials.}
\label{fig:morph}
\end{figure}
\section{Flow structures at various density ratios} \label{sec:morph}
We begin with the qualitative description on how the flow morphology changes with increasing density ratio $\Lambda$, where $\Lambda$ measures the strength of the thermal buoyancy compared to stabilization due to the stable stratification. For this study we fix $Ra_T=10^9$ and $Le=100$. In figure \ref{fig:morph} we show
the concentration and temperature fields for both regimes which emerge:

At $\Lambda=1$, when the thermal buoyancy is dominant, there is a single large-scale circulation. From the temperature field, it can be seen that the detached hot (cold) plumes travel upwards (downwards). For the chosen aspect ratio $\Gamma=1/2$ they travel over the distance of the entire cell height. At the same time, the concentration field is advected by the thermally-driven circulation. As a result, there is a well-mixed region formed in the bulk with nearly uniform concentration, whereas the concentration only changes sharply near the top and bottom boundary layers. This flow structure is similar to that in vertical convection \citep{ng2015vertical,shishkina2016thermal,wang2019non}. We thus classify this and corresponding cases into the so-called quasi-VC regime.

Strikingly, different flow structures are obtained for $\Lambda$ larger than a threshold value, which will be calculated in Section \ref{sec:height}. For example, for $\Lambda = 4$ beyond this threshold, we identify the formation of a layered structure from the concentration field. The physical process of layer formation is as follows: Initially, the detached hot plumes travel upwards. Due to the restoring force caused by stably-stratified concentration field, thermal buoyancy is not strong enough to maintain the upward-moving plumes throughout the entire domain height. Therefore, thermal plumes travel horizontally towards the middle of the cell, causing a sequence of thermal streaks as seen from the temperature field. In this case, the thermal driving leads to the vertically-stacked convection rolls. Because the concentration diffuses much slower than the heat ($Le=\kappa_T/\kappa_S=100$), a well-recognizable layered concentration field is resulted. Within each roll, the concentration is nearly uniform due to the convective mixing. At the interface between two adjacent rolls, the concentration changes sharply. For an even larger stabilization ($\Lambda=7$), even more layers initially form as compared to the case of $\Lambda=4$, in accordance with our physical explanation of the formation process.



\section{Initial layer thickness and phase space}\label{sec:height}
As shown above, a series of layers will form initially in the layered regime, and the initial layer thickness decreases with increasing strength of stabilization. What sets the initial layer thickness, or equivalently the size of the localized circulation? We will derive this initial layer thickness from an energy balance. A similar energy argument was adopted in stratified Taylor-Couette (TC) flow, which is TC flow subjected to vertical linear stratification \citep{boubnov1995stratified}. In stratified TC, spontaneous layer formation can be observed in both the low-Re \citep{boubnov1995stratified} and the high-Re \citep{oglethorpe2013spontaneous} regimes. In order to estimate the layer thickness in stratified TC, \cite{boubnov1995stratified} successfully employed the balance between the work exerted by the centripetal force and the potential energy for moving the fluid parcel in the stable stratification.

\begin{figure}
\centering
\centerline{\includegraphics[width=0.9\textwidth]{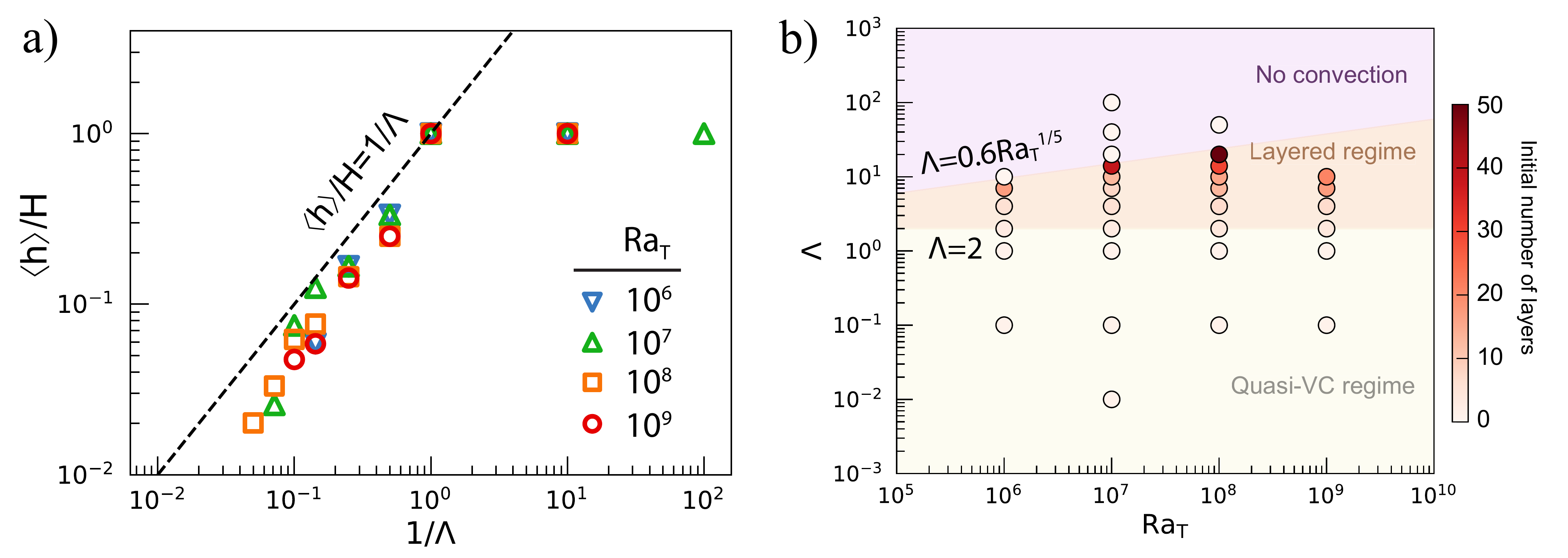}}
\caption{(a) Normalized average layer thickness $\langle h\rangle/H$ versus $1/\Lambda$ for different $Ra_T$. The black-dashed line $\langle h\rangle/H=1/\Lambda$ is derived based on the energy balance discussed in \S \ref{sec:height}. (b) Explored phase space and illustration of different flow regimes. In the quasi-VC regime, the flow resembles that in vertical convection. In the layered regime, layered structures initially emerge. The boundary between the quasi-VC and the layered regime is given by $\Lambda=2$ as derived by the energy balance in \S \ref{sec:height}. The boundary between the layered regime and the regime without convection is $\Lambda_c=0.6Ra_T^{1/5}$ as obtained already by \cite{thorpe1969effect}. The colors of the points denote the number of layers observed in the early stage of the layer formation; later these layers partly merge.}
\label{fig:height}
\end{figure}

Likewise, in laterally cooled DDC, the work for raising the fluid parcel in the stable stratification is done by the thermal buoyancy. As the fluid movement is driven by the horizontal temperature difference $\Delta_T$,  the work done by thermal buoyancy to raise the fluid parcel over distance $h$ is $(\beta_T g\Delta_T)h$. In stable linear stratification, the potential energy to bring a fluid parcel out of its equilibrium position with vertical displacement $h$ is $N^2_0h^2$, where $N_0=\sqrt{g\beta_s \Delta_s/H}$ is the buoyancy frequency. Assume that all work is converted to potential energy. This balance gives
\begin{equation}\label{eq:workdonePE}
(\beta_T g\Delta_T)h = g\beta_s \Delta_sh^2/H \qquad {\rm implying} \qquad h/H=1/\Lambda.
\end{equation}
We emphasize that this relationship is only valid for estimating the initial thickness because it assumes a linear stratification which is only the case during the initial stage.

We now check whether equation (\ref{eq:workdonePE}) is a good approximation to the initial layer thickness. We first manually count the number of layers formed in the very initial stage after layer development, which is well recognizable from the snapshots. Then the average layer thickness $\langle h \rangle$ can be estimated by dividing the cell height $H$ over the counted number. Figure \ref{fig:height}(a) shows the evaluated layer thickness $\langle h\rangle/H$ versus $1/\Lambda$ for various $Ra_T$. It can be seen that the data points generally follow the trend of $h/H=1/\Lambda$. Yet, close inspection suggests that all data points are actually below the estimated line (black dashed), consistent with previous experiments which also found that the measured initial thickness is in general less than $1/\Lambda$ \citep{chen1971stability}. There are two reasons why $h/H$ is slightly smaller than $1/\Lambda$: (i) Part of the work is not converted but dissipated which is neglected in obtaining equation (\ref{eq:workdonePE}). (ii) The buoyancy in general is smaller than $\beta_Tg\Delta_T$. Thus, equation (\ref{eq:workdonePE}) can only be seen as upper limit for the initial thickness. In addition, the layer thickness is obviously limited by the system height $H$, and indeed in figure \ref{fig:height}a it can been seen that $\langle h \rangle$ levels off at $\langle h \rangle=H$ for small $\Lambda\leq1$ or $1/\Lambda\geq1$.

We now explore the full parameter space ($\Lambda$, $Ra_T$) to identify when a single large-scale circulation forms, and when there are layers consisting of stacking localized circulations. In figure \ref{fig:height}(b), the transition boundary between the quasi-VC regime and the layered regime corresponds to the case with only two initially-formed layers, i.e., $\Lambda=2$ from equation (\ref{eq:workdonePE}). Upon increasing $\Lambda$, the number of layers increases but eventually the system reaches the motionless state when stabilization becomes dominant. This transition density ratio $\Lambda_c$ to the no convection regime was deduced previously from linear stability analysis \citep{thorpe1969effect}.

\begin{figure}
\centering
\centerline{\includegraphics[width=0.85\textwidth]{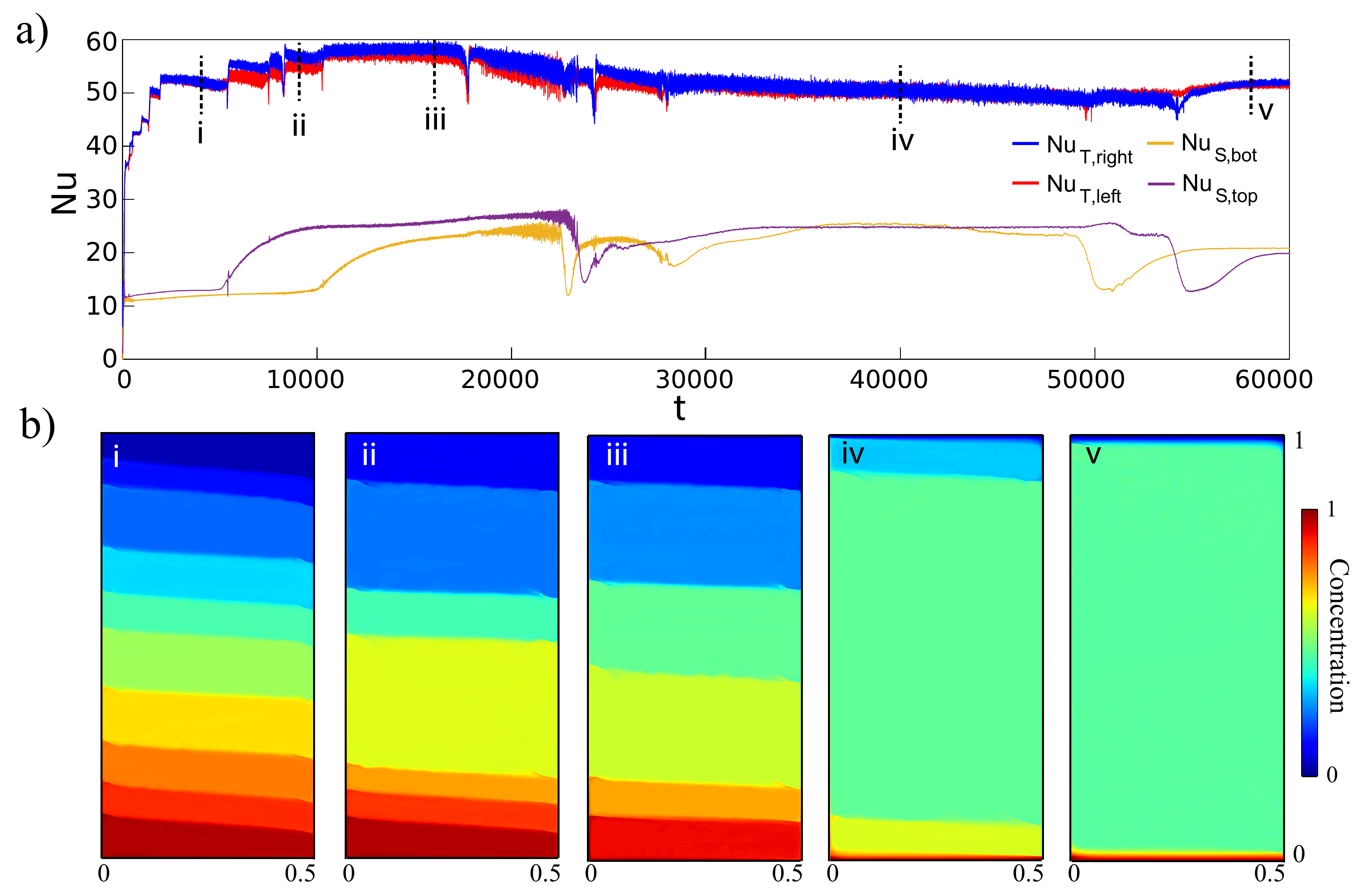}}
\caption{(a) Time series of the temperature and the concentration Nusselt number, see legend. (b) The corresponding concentration field at the marked (in Fig. \ref{fig:ts}(a)) time instants (i)-(v) are shown.}
\label{fig:ts}
\end{figure}

\section{Layer merging and its mechanism} \label{sec:mech}
We next address the merging of the layers, which sucessively occurs as time proceeds. It obviously coincides with the number of layers decreasing monotonically with time. Previous experiments had observed an eventual single roll state after successive layer merging \citep{kamakura1993experimental} for certain parameters. To study the merging process in detail, we simulated the case of $Ra_T=10^9$ and $\Lambda=7$ for $6\times10^4$ free fall time units. 

\begin{figure}
\centering
\centerline{\includegraphics[width=0.75\textwidth]{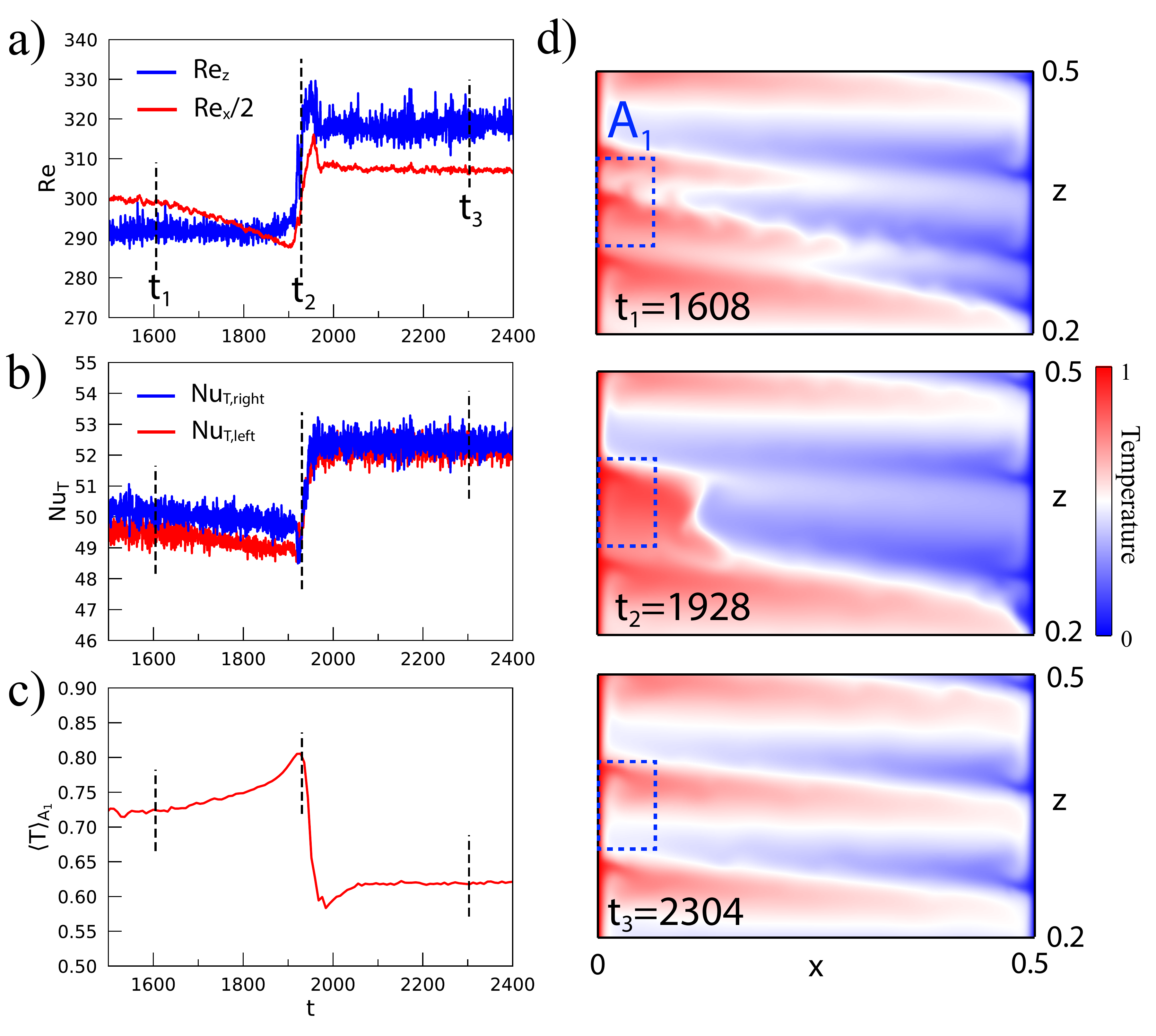}}
\caption{Time series of (a) the Reynolds number for the horizontal velocity $Re_x$ and for the vertical velocity $Re_z$. (b) Time series of the temperature Nusselt number for the left sidewall $Nu_{T,left}$, the right sidewall $Nu_{T,right}$, and (c) the averaged temperature $\langle T\rangle_{A_1}$ over the domain $A_1$. (d) Portion of the temperature snapshots at these different time instants as marked in (a)-(c). The extent of the domain $A_1$ ($0.3\le z\le0.4$ and $0\le x\le0.125$), over which $\langle T\rangle_{A_1}$ is averaged in (c), is also sketched.}
\label{fig:ts_Re}
\end{figure}

Figure \ref{fig:ts}(a) shows the time evolution of the lateral heat flux ($Nu_{T,right}$ and $Nu_{T,left}$) and the vertical solutal flux ($Nu_{S,top}$ and $Nu_{S,bot}$) for this prolonged run; the instantaneous concentration fields at different time instants are also shown in figure \ref{fig:ts}(b). Between the consecutive merging events, the system reaches a meta-stable state with fluxes fluctuating around an average value. However, just before the transition to another meta-stable state (i.e., layer merging), we observe spikes in the heat flux time series which are characteristic of layer merging. After the neighbouring layers have merged, the heat fluxes then reach another average value. In contrast, the solutal fluxes are less sensitive to the merging event as compared to the heat fluxes. The reason is that the solute diffuses a hundred times slower than the temperature, and thus it takes a longer time for the local concentration change to affect the top and bottom solutal fluxes.



The $Nu$ behaviour leads us to ask (i) Why are there spikes just before the merging of layers? (ii) What is the merging mechanism in the laterally cooled DDC? To answer these questions, we examine a particular merging event. Figure \ref{fig:ts_Re}(a) shows the vertical and horizontal Reynolds number, $Re_z$ and $Re_x$, computed by the globally-averaged horizontal and vertical velocities. Between $t_1$ and $t_2$, $Re_z$ is almost unchanged, whereas $Re_x$ progressively decreases with time during this period. The weakened horizontal velocity first explains why there is a gradual decrease in lateral heat fluxes shown in figure \ref{fig:ts_Re}(b).

To study the merging process in even more detailed, we visualize the temperature snapshots at the corresponding time instants $t_1$ and $t_2$ in figure \ref{fig:ts_Re}(d). At the moment when the layers are just about to merge ($t=t_2$), we observe that the thermal streak becomes shorter than before ($t=t_1$), reflecting the weakened local circulation within a layer. Eventually, the weakened circulation leads to an accumulation of hot fluid adjacent to the heated walls. Indeed, in figure \ref{fig:ts_Re}(c), we see that the averaged temperature over the area near the heated wall (denoted by zone $A_1$) increases gradually before $t_2$. Finally, when the hot fluid has large enough potential energy to overcome the stable stratification, two nearby layers merge into a larger one at $t_3$. A new circulation establishes after layer merging, it further carries a chunk of accumulated hot fluid near the heated wall to the opposite cold wall, and subsequently leads to the sharp increase in the heat fluxes. Our results demonstrate the complete process of layer merging in laterally cooled DDC, and further explain why there are spikes in the heat flux time series.
\section{Concluding remarks} \label{sec:conc}
In summary, inspired by the layer formation in Caf\'e Latte, we have numerically studied laterally cooled and stably density stratified double diffusive convection which is regarded as a simplified version of Caf\'e Latte. We have clearly demonstrated the layer formation, which occurs after the initial expresso injection process. We numerically explored a large range parameters, namely for the temperature Rayleigh number $10^6\leq Ra_T\leq10^9$ and for the density ratio $10^{-2}\leq\Lambda\leq10^2$, with $Le=100$.

Upon increasing strength of the stabilizing concentration gradient at a fixed lateral thermal driving, the study reveals the flow transition from the quasi-VC regime to the layered regime. In the quasi-VC regime, the flow structure is a large-scale circulation. However, in the layered regime, multiple localized circulations initially form which therefore leads to the layered structure in the concentration field. Based on the energy balance between the work done by thermal buoyancy and the potential energy to bring a fluid parcel out of its equilibrium position in stratification, we can estimate the initial layer thickness, obtaining that it roughly follows $h/H=1/\Lambda$. Such a relationship also allows us to find the boundary between quasi-VC and layered regimes, which is $\Lambda=2$.

We finally focused on merging events by running a specific case with a long time series. With sufficiently long simulations, we showed that the layered structures eventually merge into a single large-scale circulation.  The merging mechanism is that the weakened circulation within a layer leads to the accumulation of hot fluid over the hot sidewall. The hot fluid parcel at some point obtained enough buoyancy to overcome the energy barrier set by the stable stratification, and it forms a new circulation of larger size. The formation of the new circulation leads to the spikes in the heat flux time series which is characteristic for layer merging.

Until now, we have only considered the cases with fixed temperature at the heated and cooled walls. However, in many circumstances, the double diffusive convection may also be subjected to time-dependent boundary conditions, for example, abrupt temperature change caused by falling icebergs, or seasonal temperature variation. Those time-dependent forcing may have a pronounced effect on the layer formation and merging, which is worthy to be studied in the future.

\section*{Acknowledgements}
We greatly appreciate the valuable discussions with Yanshen Li and Chong Shen Ng. We acknowledge the support from an ERC-Advanced Grant under the project number $740479$. K. L. C. acknowledges Croucher Foundation for Croucher Fellowships for Postdoctoral Research. We also acknowledge PRACE for awarding us access to MareNostrum at the Barcelona Supercomputing Centre (BSC) under PRACE project number 2017174146, and DECI resource Kay based in Ireland at Irish HPC Centre. This work was also partly carried out on the national e-infrastructure of the SURFsara with the support of SURF Cooperative.


\begin{thebibliography}{34}
\expandafter\ifx\csname natexlab\endcsname\relax\def\natexlab#1{#1}\fi
\def\au#1{#1} \def\ed#1{#1} \def\yr#1{#1}\def\at#1{#1}\def\jt#1{\textit{#1}}
  \def\bt#1{#1}\def\bvol#1{\textbf{#1}} \def\vol#1{#1} \def\pg#1{#1}
  \def\publ#1{#1}\def\arxiv#1{#1}\def\org#1{#1}\def\st#1{\textit{#1}}

\bibitem[Boubnov {\em et~al.\/}(1995)Boubnov, Gledzer \&
  Hopfinger]{boubnov1995stratified}
{\sc \au{Boubnov, B.~M.}, \au{Gledzer, E.~B.} \& \au{Hopfinger, E.~J.}}
  \yr{1995}  \at{{Stratified circular Couette flow: instability and flow
  regimes}}.  \jt{J. Fluid Mech.}  \bvol{292},  \pg{333--358}.

\bibitem[Chen {\em et~al.\/}(1971)Chen, Briggs \& Wirtz]{chen1971stability}
{\sc \au{Chen, C.~F.}, \au{Briggs, D.~G.} \& \au{Wirtz, R.~A.}} \yr{1971}
  \at{Stability of thermal convection in a salinity gradient due to lateral
  heating}.  \jt{Int. J. Heat Mass Transf.}  \bvol{14}~(1),  \pg{57--65}.

\bibitem[Chen \& Chen(1997)]{chen1997salt}
{\sc \au{Chen, C.~F.} \& \au{Chen, F.}} \yr{1997}  \at{Salt-finger convection
  generated by lateral heating of a solute gradient}.  \jt{J. Fluid Mech.}
  \bvol{352},  \pg{161--176}.

\bibitem[Chong {\em et~al.\/}(2020)Chong, Yang, Yang, Verzicco \&
  Lohse]{chong2020subcritical}
{\sc \au{Chong, K.~L.}, \au{Yang, R.}, \au{Yang, Y.}, \au{Verzicco, R.} \&
  \au{Lohse, D.}} \yr{2020}  \at{Subcritical behaviour in double diffusive
  convection within the diffusive regime}.  \jt{preprint arXiv:2003.12394} .

\bibitem[Garaud(2018)]{Garaud2018a}
{\sc \au{Garaud, P.}} \yr{2018}  \at{{Double-Diffusive Convection at Low
  Prandtl Number}}.  \jt{Annu. Rev. Fluid Mech.}  \bvol{50}~(1),
  \pg{275--298}.

\bibitem[Gayen {\em et~al.\/}(2016)Gayen, Griffiths \&
  Kerr]{gayen2016simulation}
{\sc \au{Gayen, B.}, \au{Griffiths, R.~W.} \& \au{Kerr, R.~C.}} \yr{2016}
  \at{Simulation of convection at a vertical ice face dissolving into saline
  water}.  \jt{J. Fluid Mech.}  \bvol{798},  \pg{284--298}.

\bibitem[Huppert \& Turner(1980)]{huppert1980ice}
{\sc \au{Huppert, H.~E.} \& \au{Turner, J.~S.}} \yr{1980}  \at{Ice blocks
  melting into a salinity gradient}.  \jt{J. Fluid Mech.}  \bvol{100}~(2),
  \pg{367--384}.

\bibitem[Huppert \& Turner(1981)]{huppert1981double}
{\sc \au{Huppert, H.~E.} \& \au{Turner, J.~S.}} \yr{1981}  \at{Double-diffusive
  convection}.  \jt{J. Fluid Mech.}  \bvol{106},  \pg{299--329}.

\bibitem[Jacobs {\em et~al.\/}(1981)Jacobs, Huppert, Holdsworth \&
  Drewry]{jacobs1981thermohaline}
{\sc \au{Jacobs, S.~S.}, \au{Huppert, H.~E.}, \au{Holdsworth, G.} \&
  \au{Drewry, D.~J.}} \yr{1981}  \at{Thermohaline steps induced by melting of
  the erebus glacier tongue}.  \jt{J. Geophys. Res.: Oceans}  \bvol{86}~(C7),
  \pg{6547--6555}.

\bibitem[Johnson \& Kearney(2009)]{Johnson2009}
{\sc \au{Johnson, G.~C.} \& \au{Kearney, K.~A.}} \yr{2009}  \at{{Ocean climate
  change fingerprints attenuated by salt fingering?}}  \jt{Geophys. Res. Lett.}
   \bvol{36}~(21),  \pg{1--5}.

\bibitem[Kamakura \& Ozoe(1993)]{kamakura1993experimental}
{\sc \au{Kamakura, K.} \& \au{Ozoe, H.}} \yr{1993}  \at{Experimental and
  numerical analyses of double diffusive natural convection heated and cooled
  from opposing vertical walls with an initial condition of a vertically linear
  concentration gradient}.  \jt{Int. J. Heat Mass Transf.}  \bvol{36}~(8),
  \pg{2125--2134}.

\bibitem[Kelley {\em et~al.\/}(2003)Kelley, Fernando, Gargett, Tanny \&
  {\"{O}}zsoy]{Kelley2003}
{\sc \au{Kelley, D.~E.}, \au{Fernando, H. J.~S.}, \au{Gargett, A.~E.},
  \au{Tanny, J.} \& \au{{\"{O}}zsoy, E.}} \yr{2003}  \at{{The diffusive regime
  of double-diffusive convection}}.  \jt{Prog. Oceanogr.}  \bvol{56}~(3-4),
  \pg{461--481}.

\bibitem[Lee \& Hyun(1991)]{lee1991double}
{\sc \au{Lee, J.~W.} \& \au{Hyun, J.~M.}} \yr{1991}  \at{Double diffusive
  convection in a cavity under a vertical solutal gradient and a horizontal
  temperature gradient}.  \jt{Int. J. Heat Mass Transf.}  \bvol{34}~(9),
  \pg{2423--2427}.

\bibitem[Mendenhall \& Mason(1923)]{mendenhall1923stratified}
{\sc \au{Mendenhall, C.~E.} \& \au{Mason, M.}} \yr{1923}  \at{The stratified
  subsidence of fine particles}.  \jt{Proc. N. Acad. Sci.}  \bvol{9}~(6),
  \pg{199}.

\bibitem[Ng {\em et~al.\/}(2015)Ng, Ooi, Lohse \& Chung]{ng2015vertical}
{\sc \au{Ng, C.~S.}, \au{Ooi, A.}, \au{Lohse, D.} \& \au{Chung, D.}} \yr{2015}
  \at{Vertical natural convection: application of the unifying theory of
  thermal convection}.  \jt{J. Fluid Mech.}  \bvol{764},  \pg{349--361}.

\bibitem[Oglethorpe {\em et~al.\/}(2013)Oglethorpe, Caulfield \&
  Woods]{oglethorpe2013spontaneous}
{\sc \au{Oglethorpe, R. L.~F.}, \au{Caulfield, C.~P.} \& \au{Woods, A.~W.}}
  \yr{2013}  \at{{Spontaneous layering in stratified turbulent Taylor--Couette
  flow}}.  \jt{J. Fluid Mech.}  \bvol{721}.

\bibitem[Ostilla-M{\'{o}}nico {\em et~al.\/}(2015)Ostilla-M{\'{o}}nico, Yang,
  van~der Poel, Lohse \& Verzicco]{Ostilla-Monico2015}
{\sc \au{Ostilla-M{\'{o}}nico, R.}, \au{Yang, Y}, \au{van~der Poel, E.~P.},
  \au{Lohse, D.} \& \au{Verzicco, R.}} \yr{2015}  \at{{A multiple-resolution
  strategy for Direct Numerical Simulation of scalar turbulence}}.  \jt{J.
  Comput. Phys.}  \bvol{301},  \pg{308--321}.

\bibitem[van~der Poel {\em et~al.\/}(2015)van~der Poel, Ostilla-M{\'o}nico,
  Donners \& Verzicco]{van2015pencil}
{\sc \au{van~der Poel, E.~P.}, \au{Ostilla-M{\'o}nico, R.}, \au{Donners, J.} \&
  \au{Verzicco, R.}} \yr{2015}  \at{A pencil distributed finite difference code
  for strongly turbulent wall-bounded flows}.  \jt{Comput. Fluids}  \bvol{116},
   \pg{10--16}.

\bibitem[van~der Poel {\em et~al.\/}(2013)van~der Poel, Stevens \&
  Lohse]{van2013comparison}
{\sc \au{van~der Poel, E.~P.}, \au{Stevens, R. J. A.~M.} \& \au{Lohse, D.}}
  \yr{2013}  \at{{Comparison between two-and three-dimensional
  Rayleigh--B{\'e}nard convection}}.  \jt{J. Fluid Mech.}  \bvol{736},
  \pg{177--194}.

\bibitem[Radko(2013)]{Radko2013}
{\sc \au{Radko, T.}} \yr{2013} {\em {Double-Diffusive Convection}\/}.
  \publ{Cambridge University Press}.

\bibitem[Schmitt(1994)]{Schmitt1994}
{\sc \au{Schmitt, R.~W.}} \yr{1994}  \at{{Double Diffusion in Oceanography}}.
  \jt{Annu. Rev. Fluid Mech.}  \bvol{26}~(1),  \pg{255--285}.

\bibitem[Schmitt(2005)]{Schmitt2005}
{\sc \au{Schmitt, R.~W.}} \yr{2005}  \at{{Enhanced Diapycnal Mixing by Salt
  Fingers in the Thermocline of the Tropical Atlantic}}.  \jt{Science}
  \bvol{308}~(5722),  \pg{685--688}.

\bibitem[Shishkina \& Horn(2016)]{shishkina2016thermal}
{\sc \au{Shishkina, O.} \& \au{Horn, S.}} \yr{2016}  \at{Thermal convection in
  inclined cylindrical containers}.  \jt{J. Fluid Mech.}  \bvol{790}.

\bibitem[Sommer {\em et~al.\/}(2013)Sommer, Carpenter, Schmid, Lueck, Schurter
  \& W{\"{u}}est]{Sommer2013}
{\sc \au{Sommer, T.}, \au{Carpenter, J.~R.}, \au{Schmid, M.}, \au{Lueck,
  R.~G.}, \au{Schurter, M.} \& \au{W{\"{u}}est, A.}} \yr{2013}  \at{{Interface
  structure and flux laws in a natural double-diffusive layering}}.  \jt{J.
  Geophys. Res.: Oceans}  \bvol{118}~(11),  \pg{6092--6106}.

\bibitem[Tanny \& Tsinober(1988)]{tanny1988dynamics}
{\sc \au{Tanny, J.} \& \au{Tsinober, A.~B.}} \yr{1988}  \at{The dynamics and
  structure of double-diffusive layers in sidewall-heating experiments}.
  \jt{J. Fluid Mech.}  \bvol{196},  \pg{135--156}.

\bibitem[Thorpe {\em et~al.\/}(1969)Thorpe, Hutt \& Soulsby]{thorpe1969effect}
{\sc \au{Thorpe, S.~A.}, \au{Hutt, P.~K.} \& \au{Soulsby, R.}} \yr{1969}
  \at{The effect of horizontal gradients on thermohaline convection}.  \jt{J.
  Fluid Mech.}  \bvol{38}~(2),  \pg{375--400}.

\bibitem[Timmermans {\em et~al.\/}(2008)Timmermans, Toole, Krishfield \&
  Winsor]{Timmermans2008}
{\sc \au{Timmermans, M.~L.}, \au{Toole, J.}, \au{Krishfield, R.} \& \au{Winsor,
  P.}} \yr{2008}  \at{{Ice Tethered Profiler observations of the double
  diffusive staircase in the Canada Basin thermocline}}.  \jt{J. Geophys. Res.:
  Oceans}  \bvol{113},  \pg{0--02}.

\bibitem[Turner(1974)]{Turner1974}
{\sc \au{Turner, J.~S.}} \yr{1974}  \at{{Double diffusive phenomena}}.
  \jt{Annu. Rev. Fluid Mech.}  \bvol{6}~(1),  \pg{37--56}.

\bibitem[Verzicco \& Orlandi(1996)]{verzicco1996finite}
{\sc \au{Verzicco, R.} \& \au{Orlandi, P.}} \yr{1996}  \at{A finite-difference
  scheme for three-dimensional incompressible flows in cylindrical
  coordinates}.  \jt{J. Comp. Phys.}  \bvol{123}~(2),  \pg{402--414}.

\bibitem[Wang {\em et~al.\/}(2019)Wang, Xia, Yan, Sun \& Wan]{wang2019non}
{\sc \au{Wang, Q.}, \au{Xia, S.-N.}, \au{Yan, R.}, \au{Sun, D.-J.} \& \au{Wan,
  Z.-H.}} \yr{2019}  \at{{Non-Oberbeck-Boussinesq effects due to large
  temperature differences in a differentially heated square cavity filled with
  air}}.  \jt{Int. J. Heat Mass Transf.}  \bvol{128},  \pg{479--491}.

\bibitem[Wirtz {\em et~al.\/}(1972)Wirtz, Briggs \& Chen]{wirtz1972physical}
{\sc \au{Wirtz, R.~A.}, \au{Briggs, D.~G.} \& \au{Chen, C.~F.}} \yr{1972}
  \at{Physical and numerical experiments on layered convection in a
  density-stratified fluid}.  \jt{Geophys. Fluid Dyn.}  \bvol{3}~(1),
  \pg{265--288}.

\bibitem[Xue {\em et~al.\/}(2017)Xue, Khodaparast, Zhu, Nunes, Kim \&
  Stone]{xue2017laboratory}
{\sc \au{Xue, N.}, \au{Khodaparast, S.}, \au{Zhu, L.}, \au{Nunes, J.~K.},
  \au{Kim, H.} \& \au{Stone, H.~A.}} \yr{2017}  \at{{Laboratory layered
  Latte}}.  \jt{Nat. Commun.}  \bvol{8}~(1),  \pg{1960}.

\bibitem[Yang {\em et~al.\/}(2015)Yang, Verzicco \& Lohse]{Yang2015}
{\sc \au{Yang, Y.}, \au{Verzicco, R.} \& \au{Lohse, D.}} \yr{2015}  \at{{From
  convection rolls to finger convection in double-diffusive turbulence}}.
  \jt{Proc. N. Acad. Sci.}  \bvol{113}~(1),  \pg{69--73}.

\bibitem[Yang {\em et~al.\/}(2019)Yang, Verzicco \& Lohse]{yang2019multiple}
{\sc \au{Yang, Y.}, \au{Verzicco, R.} \& \au{Lohse, D.}} \yr{2019}
  \at{{Multiple equilibria in fingering double diffusive convection
  turbulence}}.  \jt{arXiv preprint arXiv:1904.09519} .

\end{thebibliography}
\end{document}